\documentclass[pra,twocolumn,showpacs,preprintnumbers,superscriptaddress]
{revtex4}

\usepackage{times}
\usepackage{bm}
\usepackage{float}
\usepackage{graphicx}
\usepackage{amsbsy}
\usepackage{amsmath}
\usepackage{amsfonts}
\usepackage{amsthm}
\begin{document}

\theoremstyle{plain}
\newtheorem{theorem}{Theorem}
\newtheorem{lemma}[theorem]{Lemma}
\newtheorem{corollary}[theorem]{Corollary}
\newtheorem{proposition}[theorem]{Proposition}
\newtheorem{conjecture}[theorem]{Conjecture}

\theoremstyle{definition}
\newtheorem{definition}[theorem]{Definition}

\theoremstyle{remark}
\newtheorem*{remark}{Remark}
\newtheorem{example}{Example}
\title{Coherence based inequality for the discrimination of three-qubit GHZ and W class}
\author{Pranav Kairon, Mukhtiyar Singh, Satyabrata Adhikari}
\email{pranav_2k17ep54@dtu.ac.in,mukhtiyarsingh@dtu.ac.in, satyabrata@dtu.ac.in} \affiliation{Delhi Technological
University, Delhi-110042, Delhi, India}

\begin{abstract}
Quantum coherence and entanglement orignate from the superposition principle. We derive a rigorous relation between the ${l_1}$-norm of coherence and concurrence, in that we show that the former is always greater than the latter. This result highlights the hierarchical relationship between coherence and concurrence, and proves coherence to be a fundamental and ubiquitous resource. We derive an analogous form of monogamy inequality which is based on the partial coherence of the reduced two-qubit and reduced single qubit of the particular class of three-qubit state. Moreover, we provide coherence based inequality for the classification of GHZ class and W class of three-qubit states. Finally, we provide theoretical discussion for the possible implementation of the scheme in an experiment.
\end{abstract}
 \maketitle

\section{Introduction}

Quantum coherence and entanglement are arguably the most significant phenomenon appear in quantum mechanics that mark the  departure from classical mechanics. Entanglement has no classical analogue but unlike this purely quantum mechanical phenomenon, coherence is a familiar event in optics. Although, quantum theory of coherence forms the foundation of the study and manipulation of optical coherence phenomenon, there is a significant difference between the two which has been studied and demonstrated through the multi-point correlation functions \cite{sudarshan1963equivalence} and the phase space representation of quantum mechanics \cite{glauber1963coherent}. This works well to distinguish between the classical and quantum phenomenon, but fails to quantify the amount of coherence present in a given system. To overcome this caveat, recently the resource theory of coherence was formulated by Buamgratz et al. \cite{baumgratz2014quantifying}. It provides a quantum information theoretic framework to quantify and manipulate coherence levels in a system. The proper measure of coherence is needed to quantify the amount of coherence present in a quantum system. To probe this, they prescribed some postulates that an ideal measure of quantum coherence must satisfy.  This has prompted various applications of quantum coherence to variegated fields such as thermodynamics \cite{engel2007evidence}, quantum metrology and sensing\cite{giorda2017coherence}, one-way quantum computing \cite{raussendorf2001one} and quantum biology \cite{narasimhachar2015low}. Quantum information protocols such as quantum secret sharing \cite{hillery1999quantum}, quantum private query\cite{gao2019quantum} also exploit quantum coherence as a resource. Some of the important works to formulate an efficient resource theory of coherence are dilineated in an extensive review\cite{streltsov2017colloquium}.\\
Entanglement and coherence both arise from the superposition principle of quantum physics and are considered to be the key concepts for quantum technologies. Unlike entanglement, the amount of coherence depends on basis and thus the application of local unitary transformations on the quantum system may enhance the amount of coherence present in a system. In \cite{yao2015quantum}, a heirarchial relationship among quantum coherence, discord and entanglement is presented, which proves coherence as a fundamental manifestation of quantum correlations.
The motivation of this work lies in the following facts: (i) Entanglement serve as a vital resource in various quantum information processing tasks  such as teleportation, super-dense coding etc. But entanglement is quite expensive and difficult to prepare in comparison to other resources such as discord, coherence. Hence it is imperative to determine a hierarchical relation between entanglement and coherence. (ii)  The well known monogamy inequality has been derived for quantum entanglement \cite{coffman} and discord \cite{modi}. This motivate us to derive an analogous monogamy inequality based on the partial coherence of the two-qubit reduced state and single qubit reduced state. (iii) There exist various methods based on entanglement by which GHZ class and W class can be distinguished but there does not exist any method based on coherence by which we can distinguish GHZ class and W class. This is the driving force for the derivation of coherence based inequality that may help to discriminate GHZ class and W class.\\
This paper is organized as follows: In Sec. II, we have studied the relationship between coherence and concurrence of an arbitrary two-qubit state. In Sec. III, we have derived an inequality based on coherence that is analogous to concurrence based monogamy inequality. Furthermore, we have constructed coherence based inequality that may discriminate between GHZ class and W class. We conclude in Sec. IV.
\section{Hierarchical relationship between coherence and Concurrence of an arbitrary two-qubit state}
Superposition principle manifests itself in two ways in quantum mechanics: quantum coherence and quantum entanglement. Zhao et. al. \cite{zhao2020coherence} have studied the relationship between coherence concurrence and negativity for the particular class of two-qubit bipartite quantum states. The complementarity relation between the entanglement of formation and quantum coherence has been obtained by Pan et.al \cite{pan2017complementarity}. Stretslov et al. \cite{streltsov2015measuring} have shown that there exist incoherent operations by which coherence can be converted to entanglement. A generalized process is given in \cite{killoran2016converting}, which shows general scheme to produce entanglement using nonclassicality as a resource. In this section, we study the hierarchical relationship between the concurrence of an arbitrary two-qubit bipartite quantum state and its ${l_1}$-norm of coherence.\\
The ${l_1}$-norm of quantum coherence is defined as summation of modulus of the off-diagonal terms of given quantum state described by the two-qubit density matrix $\rho$ \cite{baumgratz2014quantifying},
\begin{equation}
C_{l_{1}}(\rho)= \sum_{i,j,i\neq j}|\rho_{ij}|,~~~~ i,j=1,2,3,4
\label{eq: 1}
\end{equation}
For any two-qubit density matrix $\rho$, concurrence can be defined as \cite{wootters}
\begin{equation}
C(\rho)= max{\left\lbrace 0,\sqrt{\lambda_{1}}-\sqrt{\lambda_{2}}-\sqrt{\lambda_{3}}-\sqrt{\lambda_{4}}\right\rbrace} \label{eq2}
\end{equation}
where $\lambda_{1}\geq \lambda_{2}\geq \lambda_{3}\geq \lambda_{4}$ denoting the eigenvalues of the matrix $\rho\bar{\rho}$, where $\bar{\rho}=(\sigma_{y}\text{\ensuremath{\otimes}}\sigma_{y})\rho^\text{\textasteriskcentered}(\sigma_{y}\text{\ensuremath{\otimes}}\sigma_{y})$ referred to as the spin flipped density matrix and $\sigma_{y}=-i|0\rangle\langle 1|+i|1\rangle\langle 0|$, represent the Pauli matrix.\\
\textbf{Theorem 1}: \textit{For any general two-qubit density matrix $\rho$, $l_{1}$-norm of quantum coherence $(C_{l_{1}}(\rho))$ is always greater than or equal to concurrence of $\rho$ $(C(\rho))$. Mathematically, it can be expressed as}
\begin{eqnarray}
C_{l_{1}}(\rho) \geq C(\rho)
\label{inequality1231}
\end{eqnarray}
\textbf{Proof:} Let us consider an arbitrary two-qubit quantum state described by the density operator $\rho$. If the state $\rho$ is separable then the inequality (\ref{inequality1231}) holds trivially. Therefore, it is sufficient to prove the inequality (\ref{inequality1231}) when the state described by the density operator $\rho$ is entangled. Thus, the concurrence of the entangled state $\rho$ is given by
\begin{eqnarray}
&&C(\rho)=\sqrt{\lambda_{1}}-\sqrt{\lambda_{2}}-\sqrt{\lambda_{3}}-\sqrt{\lambda_{4}}\nonumber\\&&
\Rightarrow C(\rho)\leq\sqrt{\lambda_{1}}=S_{max}(\rho\bar{\rho})\leq S_{max}(\rho)S_{max}(\bar{\rho})
\label{eq3}
\end{eqnarray}
where $S_{max}(\rho\tilde{\rho} )$ denotes the maximum singular value of the matrix $\rho\tilde{\rho}$. The last inequality follows from \cite{horn}.\\
The inequality (\ref{eq3}) can be further simplified by using the result $S_{max}(\rho)\leq\left\| \rho\right\|_{2}$, where $\|\rho\|_{2}=Tr(\rho^{2})$ \cite{zhan}. Then the inequality (\ref{eq3}) reduces to
\begin{eqnarray}
C(\rho)&\leq& \left\| \rho\right\|_{2}.S_{max}(\tilde{\rho})\nonumber\\&\leq&
\left\| \rho\right\|_{1}.S_{max}(\tilde{\rho})
\label{eq4}
\end{eqnarray}
The last inequality follows from the fact that $\left\| \rho\right\|_{p}\leq\left\| \rho\right\|_{q}$ for $0<q\leq p$ \cite{horn}. The relation between $\left\| \rho\right\|_{1}$ and $C_{l_{1}}(\rho)$ is given by (For details, see Appendix-A)
\begin{eqnarray}
\left\| \rho\right\|_{1}\leq C_{l_{1}}(\rho)
\label{eq5}
\end{eqnarray}
Combining the results (\ref{eq4} and (\ref{eq5}), we get
\begin{eqnarray}
C(\rho)\leq C_{l_{1}}(\rho).S_{max}(\tilde{\rho}) ,\\
\Rightarrow\frac{C(\rho)}{C_{l_{1}}(\rho)}\leq S_{max}(\tilde{\rho})\leq 1
\label{eq6}
\end{eqnarray}
Thus, we get the required result
\begin{eqnarray}
C(\rho)\leq C_{l_{1}}(\rho)
\label{eq7}
\end{eqnarray}
Hence the theorem is proved.
\begin{figure}
\begin{centering}
\includegraphics[scale=0.45]{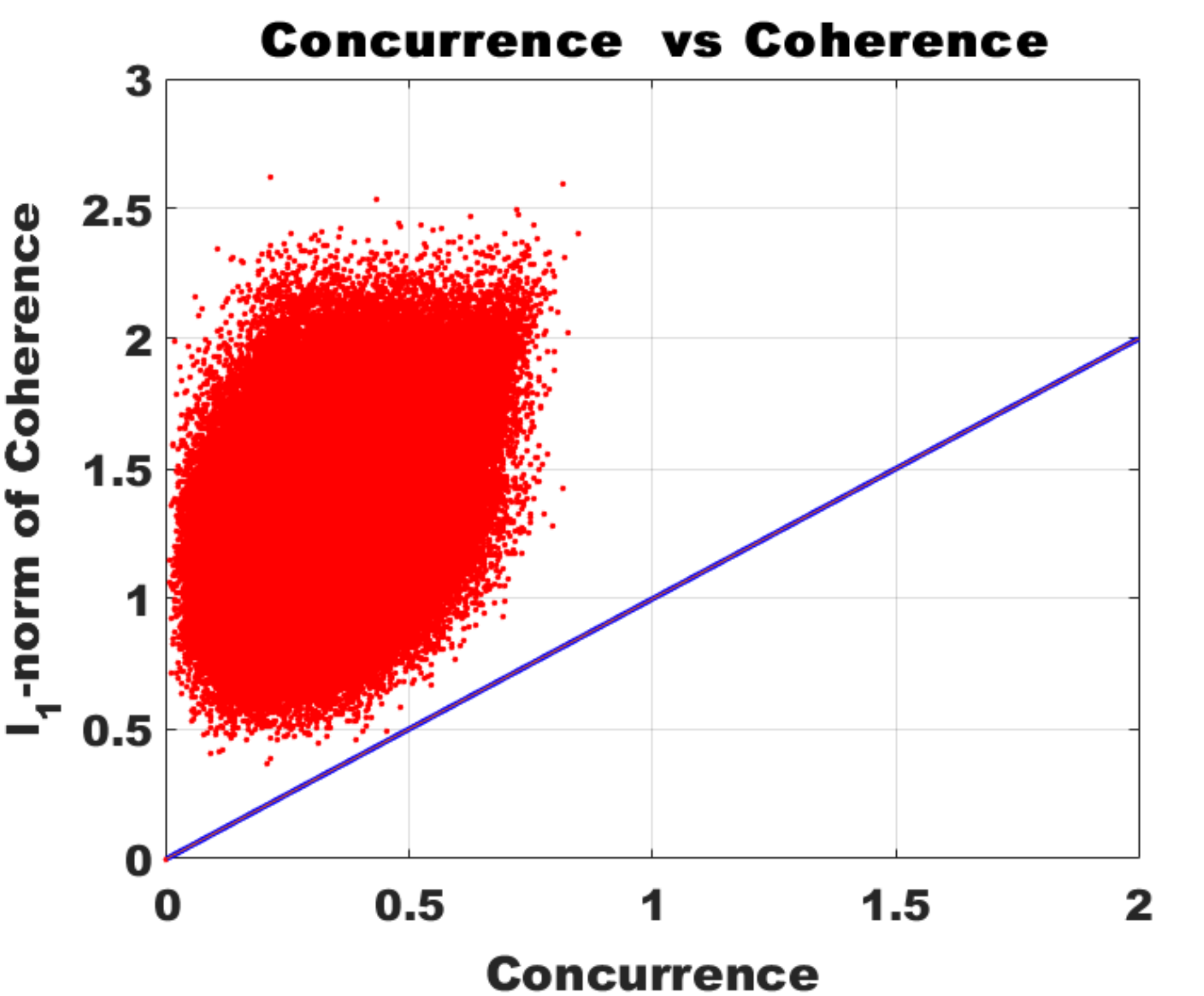}
\par\end{centering}
\caption{\label{fig:circuit}Although we have provided general proof, our results are backed up by the numerical analysis of $10^{6}$ haar-randomly generated density matrices. The blue line is $y=x$ and all the points lie in the region $y\geq x$ which corroborates Theorem 1}
\end{figure}

\section{Coherence based inequality analogous to the concurrence based inequality derived by coffman et.al.}
In this section, we will derive an inequality that provide us the upper bound of the sum of the coherences of the reduced two-qubit of the particular class of pure three-qubit system. This inequality is derived in the spirit of the seminal work by Coffman et.al. \cite{coffman}.\\
For a pure three-qubit state $|\psi\rangle_{ABC}$, Coffman et.al. have derived an inequality based on the concurrence between the qubits $A$ and $B$ and the qubits $A$ and $C$. Mathematically, this inequality can be expressed as \cite{coffman}
\begin{eqnarray}
C_{AB}^{2}+C_{AC}^{2}\leq C_{A(BC)}^{2}
\label{ckw}
\end{eqnarray}
where $C_{AB}$ and $C_{AC}$ denote the partial concurrences that measures the amount of entanglement in the reduced two-qubit mixed state described by the density operators $\rho_{AB}$ and $\rho_{AC}$ respectively of the pure three-qubit state $|\psi\rangle_{ABC}$. $C_{A(BC)}$ denote the concurrence between subsystems $A$ and $BC$. The inequality (\ref{ckw}) is also known as monogamous inequality. Here, our aim is to construct the coherence based inequality analogous to the concurrence based inequality (\ref{ckw}).\\
Any state vector in a three-qubit system can be spanned by eight computational basis vectors and thus we require seven parameters to characterize a general three-qubit quantum state. But Acin et al. \cite{acin1} have deduced the canonical form of three-qubit state and shown that it can be represented by five parameters only. This idea was later on proved to be true for multipartite states by Cateret et. al.\cite{carteret2000multipartite}.\\
The canonical form of three-qubit state can be expressed as \cite{acin1}
\begin{eqnarray}
\left|\psi\right\rangle_{ABC}^{(\theta)} &=&\lambda_{0}\left|000\right\rangle  +\lambda_{1}e^{i\theta}\left|100\right\rangle +\lambda_{2}\left|101\right\rangle +\lambda_{3}\left|110\right\rangle\nonumber\\&+&\lambda_{4}\left|111\right\rangle
\label{eq8}
\end{eqnarray}
where the state parameters $\lambda_{i}\geq 0, (i=0,1,2,3,4)$ and the phase factor $0\leq \theta \leq \pi$.\\
The normalization condition gives
\begin{eqnarray}
\lambda_{0}^{2}+\lambda_{1}^{2}+\lambda_{2}^{2}+\lambda_{3}^{2}+\lambda_{4}^{2}=1
\label{normalization1}
\end{eqnarray}
If there is no phase factor i.e. $\theta=0$ then the three-qubit state (\ref{eq8}) reduces to a particular class, which is represented by
\begin{eqnarray}
\left|\psi\right\rangle_{ABC}^{(0)} &=&\lambda_{0}\left|000\right\rangle  +\lambda_{1}\left|100\right\rangle +\lambda_{2}\left|101\right\rangle +\lambda_{3}\left|110\right\rangle\nonumber\\&+&\lambda_{4}\left|111\right\rangle
\label{particularclass}
\end{eqnarray}
The relations between the state parameters and the partial concurrences of the pure three-qubit state $|\psi\rangle_{ABC}^{(0)}$ are invariants under local unitary transformation and these invariant relations are given by \cite{torun}
\begin{eqnarray}
&&C_{AB}=2\lambda_{0}\lambda_{3}\nonumber\\&&
C_{AC}=2\lambda_{0}\lambda_{2}
\label{eq9}
\end{eqnarray}
$l_{1}$ norm of coherence for the reduced two-qubit states and single qubit state is described by the density operators $\rho_{AB}$, $\rho_{AC}$ and $\rho_{A}$ respectively of the pure three-qubit state $|\psi\rangle_{ABC}^{(0)}$ are given by
\begin{eqnarray}
C_{l_{1}}(\rho_{AB})=2(\lambda_{0}\lambda_{1}+\lambda_{0}\lambda_{3}+\lambda_{1}\lambda_{3}+\lambda_{2}\lambda_{4})
\label{cohab}
\end{eqnarray}
\begin{eqnarray}
C_{l_{1}}(\rho_{AC})=2(\lambda_{0}\lambda_{1}+\lambda_{0}\lambda_{2}+\lambda_{1}\lambda_{2}+\lambda_{3}\lambda_{4})
\label{cohac}
\end{eqnarray}
\begin{eqnarray}
C_{l_{1}}(\rho_{A})=2\lambda_{0}\lambda_{1}
\label{coha}
\end{eqnarray}
Squaring (\ref{cohab})and (\ref{cohac}) and then adding, we get
\begin{eqnarray}
C_{l_{1}}^{2}(\rho_{AB})+C_{l_{1}}^{2}(\rho_{AC})&=&4[(\lambda_{0}(\lambda_{1}+\lambda_{3})+
\lambda_{1}\lambda_{3}+\lambda_{2}\lambda_{4})^{2}\nonumber\\&+&(\lambda_{0}(\lambda_{1}+\lambda_{2})+
\lambda_{1}\lambda_{2}+\lambda_{3}\lambda_{4})^{2}]\nonumber\\&=&
C_{AB}^{2}+C_{AC}^{2}+2C_{l_{1}}^{2}(\rho_{A})\nonumber\\&+&(\textrm{Sum of positive terms})
\label{cohabac}
\end{eqnarray}
The equation (\ref{cohabac}) can also be expressed as
\begin{eqnarray}
C_{l_{1}}^{2}(\rho_{AB})+C_{l_{1}}^{2}(\rho_{AC})-2C_{l_{1}}^{2}(\rho_{A})&=&\textrm{Sum of all finite positive}
\nonumber\\&& \textrm{ numbers} \nonumber\\&\geq& 0
\label{cohabac1}
\end{eqnarray}
Hence, the required inequality is given by
\begin{eqnarray}
C_{l_{1}}^{2}(\rho_{AB})+C_{l_{1}}^{2}(\rho_{AC})\geq 2C_{l_{1}}^{2}(\rho_{A})
\label{coherenceineq}
\end{eqnarray}
The inequality (\ref{coherenceineq}) can be considered as an analogous form of the concurrence based monogamous inequality (\ref{ckw}). The inequality (\ref{coherenceineq}) holds for the particular type of large class of pure three-qubit state $\left|\psi\right\rangle_{ABC}^{(0)}$.

\section{Few inequalities based on coherence}
In this section, we will derive the coherence based inequality that may be used to discriminate GHZ class and W class of pure three-qubit state. Then we also characterize GHZ class of state based on few coherence based inequalities.
\subsection{Discrimination of pure three-qubit GHZ and W class using coherence based inequality}
GHZ class and W class represents two genuine entangled class of three-qubit pure state, which are inequivalent under stochastic local operation and classical communication (SLOCC). The amount of entanglement in three-qubit state belong to GHZ class can be quantified by the non-zero value of the three tangle denoted by $\tau$. For any pure three-qubit state $|\psi\rangle_{ABC}$, it can be defined as residual entanglement \cite{coffman}
\begin{eqnarray}
\tau= C_{A(BC)}^{2}-C_{AB}^{2}-C_{AC}^{2}
\label{tangledef}
\end{eqnarray}
The tangle for the state $|\psi\rangle_{ABC}^{\theta}$ can be calculated as
\begin{eqnarray}
\tau_{|\psi\rangle_{ABC}^{\theta}}= 4\lambda_{0}^{2}\lambda_{4}^{2}
\label{tanglecal}
\end{eqnarray}
If the state parameters $\lambda_{0}$ and $\lambda_{4}$ are non-zero then the state belong to GHZ class can be expressed in the following form
\begin{eqnarray}
\left|\psi\right\rangle_{GHZ}^{(0)} &=&\lambda_{0}\left|000\right\rangle  +\lambda_{1}\left|100\right\rangle +\lambda_{2}\left|101\right\rangle +\lambda_{3}\left|110\right\rangle\nonumber\\&+&\lambda_{4}\left|111\right\rangle
\label{wclass}
\end{eqnarray}
The three tangle vanishes for W class of states. Therefore, either $\lambda_{0}=0$ or $\lambda_{4}=0$ for W class of states. From (\ref{eq9}), we can observe that if we take $\lambda_{0}=0$ then $C_{AB}=C_{AC}=0$. Therefore, it would be advisable to take $\lambda_{4}=0$ for W class of state and it is expressed in the form given by
\begin{eqnarray}
\left|\psi\right\rangle_{W}^{(0)} &=&\lambda_{0}\left|000\right\rangle  +\lambda_{1}\left|100\right\rangle +\lambda_{2}\left|101\right\rangle +\lambda_{3}\left|110\right\rangle
\label{wclass}
\end{eqnarray}
Thus, it may appear that tangle can be a suitable candidate for the classification of GHZ class and W class. But since tangle remain zero for three-qubit biseparable and separable classes of states so it is not possible to conclude that the given class represent a W class if the tangle is zero. Thus, we derive here coherence based inequality that may be used to classify pure three-qubit GHZ and W class of states.\\
Let us first recall (\ref{particularclass}), which represent the canonical form of pure three-qubit state $|\psi\rangle_{ABC}^{(0)}$. To start with the derivation of the inequality, let us consider the expression $C_{l_{1}}(\rho_{AB})-C_{l_{1}}(\rho_{AC})$ which is given by
\begin{eqnarray}
C_{l_{1}}(\rho_{AB})-C_{l_{1}}(\rho_{AC})=2(\lambda_{3}-\lambda_{2})(\lambda_{0}+\lambda_{1}-\lambda_{4})
\label{exp1}
\end{eqnarray}
Now we can consider two cases based on the sign of the expression $(\lambda_{3}-\lambda_{2})$.\\
\textbf{Case-I:} If $\lambda_{3}-\lambda_{2}\geq 0$ then we can observe the following points considered below:\\
(i) $C_{l_{1}}(\rho_{AB})-C_{l_{1}}(\rho_{AC})\geq 0$, for every state belong to W class given by $\left|\psi\right\rangle_{W}^{(0)}$.\\
(ii) $C_{l_{1}}(\rho_{AB})-C_{l_{1}}(\rho_{AC})< 0$, for at least one state belong to GHZ class given by $\left|\psi\right\rangle_{GHZ}^{(0)}$.\\
\textbf{Case-II:} If $\lambda_{3}-\lambda_{2} < 0$ then we have the following:\\
(i) $C_{l_{1}}(\rho_{AB})-C_{l_{1}}(\rho_{AC})< 0$, for every state belong to W class given by $\left|\psi\right\rangle_{W}^{(0)}$.\\
(ii) $C_{l_{1}}(\rho_{AB})-C_{l_{1}}(\rho_{AC})\geq 0$, for at least one state belong to GHZ class given by $\left|\psi\right\rangle_{GHZ}^{(0)}$.\\
\subsection{Few results on the characterization of GHZ class}
We discuss here few results which will be applicable only for the states belong to GHZ class.\\
\textbf{Result-1:} If the state belong to GHZ class and choose the state parameters $\lambda_{0}$, $\lambda_{1}$ and $\lambda_{4}$ in such a way so that $\lambda_{0}+\lambda_{1}-\lambda_{4} <0$ holds then
\begin{eqnarray}
C_{AB}+C_{AC}<2 C_{l_{1}}(\rho_{AC})
\label{ineq1}
\end{eqnarray}
\textbf{Proof:} Let us consider any state belongs to $\left|\psi\right\rangle_{GHZ}^{(0)}$. Then using the result in Theorem-1, the sum of the partial concurrences $C_{AB}$ and $C_{AC}$ can be expressed as
\begin{eqnarray}
C_{AB}+C_{AC}&\leq& C_{l_{1}}(\rho_{AB})+C_{l_{1}}(\rho_{AC})\nonumber\\&<&
2C_{l_{1}}(\rho_{AC})
\label{ineq1}
\end{eqnarray}
If $\lambda_{0}+\lambda_{1}-\lambda_{4} <0$ holds for the state belong to GHZ class $\left|\psi\right\rangle_{GHZ}^{(0)}$ then we have $C_{l_{1}}(\rho_{AB})<C_{l_{1}}(\rho_{AC})$ and we achieved The last inequality. Hence proved.\\

\textbf{Result-2:} If the state belong to GHZ class and the inequality $\lambda_{0}+\lambda_{1}-\lambda_{4} <0$ holds for some state parameters $\lambda_{0}$, $\lambda_{1}$ and $\lambda_{4}$ then
\begin{eqnarray}
C_{l_{1}}(\rho_{A})<C_{l_{1}}(\rho_{AC})
\label{ineq2}
\end{eqnarray}
\textbf{Proof:} Recalling (\ref{coherenceineq}) and re-expressing it as
\begin{eqnarray}
\frac{C_{l_{1}}^{2}(\rho_{AB})+C_{l_{1}}^{2}(\rho_{AC})}{2}\geq C_{l_{1}}^{2}(\rho_{A})
\label{coherenceineq1}
\end{eqnarray}
Using AM-GM inequality on $C_{l_{1}}^{2}(\rho_{AB})$ and $C_{l_{1}}^{2}(\rho_{AC})$, we get
\begin{eqnarray}
\frac{C_{l_{1}}^{2}(\rho_{AB})+C_{l_{1}}^{2}(\rho_{AC})}{2}\geq  C_{l_{1}}(\rho_{AB})C_{l_{1}}(\rho_{AC})
\label{amgm}
\end{eqnarray}
From (\ref{coherenceineq1}) and (\ref{amgm}), it is not clear that whether $C_{l_{1}}(\rho_{AB})C_{l_{1}}(\rho_{AC})-C_{l_{1}}^{2}(\rho_{A})\geq 0$ or $C_{l_{1}}(\rho_{AB})C_{l_{1}}(\rho_{AC})-C_{l_{1}}^{2}(\rho_{A})< 0$ holds. To investigate this, let us express the value of the expression $C_{l_{1}}(\rho_{AB})C_{l_{1}}(\rho_{AC})-C_{l_{1}}^{2}(\rho_{A})$ in terms of the state parameters. We have
\begin{eqnarray}
C_{l_{1}}(\rho_{AB})C_{l_{1}}(\rho_{AC})-C_{l_{1}}^{2}(\rho_{A})&=&4\lambda_{0}\lambda_{1}\lambda_{2}(\lambda_{0}+\lambda_{1})\nonumber\\&+&
4\lambda_{3}(\lambda_{0}+\lambda_{1}) (\lambda_{0}\lambda_{1}+\nonumber\\&&\lambda_{0}\lambda_{2}+\lambda_{1}\lambda_{2})
\label{exp12}
\end{eqnarray}
Since all $\lambda_{i}\geq 0$ so we get
\begin{eqnarray}
C_{l_{1}}(\rho_{AB})C_{l_{1}}(\rho_{AC})\geq C_{l_{1}}^{2}(\rho_{A})
\label{ineq12}
\end{eqnarray}
If $\lambda_{0}+\lambda_{1}-\lambda_{4} <0$ holds for the state belong to GHZ class $\left|\psi\right\rangle_{GHZ}^{(0)}$ then we have $C_{l_{1}}(\rho_{AB})<C_{l_{1}}(\rho_{AC})$ and using the result given in (\ref{ineq12}), we get
\begin{eqnarray}
&&C_{l_{1}}^{2}(\rho_{A})\leq C_{l_{1}}^{2}(\rho_{AC})\nonumber\\&&
\Rightarrow C_{l_{1}}(\rho_{A})< C_{l_{1}}(\rho_{AC})
\label{result2}
\end{eqnarray}
Hence proved.
\subsection{Experimental realization of the inequality $\lambda_{0}+\lambda_{1}-\lambda_{4}<0$}
In the previous sections, we have seen that the inequality $\lambda_{0}+\lambda_{1}-\lambda_{4}<0$ play an important role in the discrimination of GHZ class and W class and also take part in the characterization of GHZ class. By seeing its importance in the characterization and classification problem, we provide here the theoretical prescription of the experimental realization of the inequality $\lambda_{0}+\lambda_{1}-\lambda_{4}<0$.\\
Multiplying by $\lambda_{0}> 0$ both sides of the inequality $\lambda_{0}+\lambda_{1}-\lambda_{4}<0$, we get
\begin{eqnarray}
&&\lambda_{0}^{2}+\lambda_{0}\lambda_{1}-\lambda_{0}\lambda_{4}<0\nonumber\\&&
\Rightarrow \sqrt{\tau}\geq2\lambda_{0}(\lambda_{0}+\lambda_{1})\nonumber\\&&
\Rightarrow \left\langle O\right\rangle_{|\psi\rangle_{ABC}^{(0)}} > \left\langle O_{1}\right\rangle_{|\psi\rangle_{ABC}^{(0)}}+\left\langle O_{2}\right\rangle_{|\psi\rangle_{ABC}^{(0)}}
\label{eq16}
\end{eqnarray}
where the operators $O$, $O_{1}$ and $O_{2}$ can be decomposed in terms of Pauli matrices as
\begin{eqnarray}
O = 2(\sigma_{x}\otimes \sigma_{x} \otimes \sigma_{x})
\label{operator1}
\end{eqnarray}
\begin{eqnarray}
O_{1} = 2(\sigma_{x}\otimes \sigma_{z} \otimes \sigma_{z})
\label{operator2}
\end{eqnarray}
\begin{eqnarray}
O_{2} = \frac{1}{4} [(I+\sigma_{z})\otimes (I+\sigma_{z}) \otimes (I+\sigma_{z})]
\label{operator3}
\end{eqnarray}
The expectation value of the operators $\left\langle O\right\rangle_{|\psi\rangle_{ABC}^{(0)}},\left\langle O_{1}\right\rangle_{|\psi\rangle_{ABC}^{(0)}},\left\langle O_{2}\right\rangle_{|\psi\rangle_{ABC}^{(0)}}$ pave a way for the possible implementation of the technique to distinguish GHZ class and W class. A classification protocol introduced in \cite{datta2018distinguishing} has been implemented on an NMR based quantum information processor by Singh et al. \cite{singh2018experimental}. Thus we believe that the classification scheme studied in this work may be implemented in NMR based experiment

\section{Conclusion}
To summarize, we have illustrated a rigorous proof that ${l_1}$-norm of coherence is greater than concurrence for a general two qubit system. In the context of partial concurrence based monogamy inequality, we have designed an analogous form of monogamy inequality based on partial coherence of the special large class of three-qubit pure state.  We have also derived the partial coherence based inequality to distinguish between GHZ and W-class of states and further we have characterized three-qubit GHZ class on the basis of the constructed inequality. We have corroborated our theoretical efforts by providing an experimental scheme to implement our proposal. We believe that this work may deepen our understanding of coherence as a resource and may provide us with better insights to manifest quantum technologies.

\section*{Appendix-A:To prove $\|\rho\|_{1}\leq C_{l_{1}}(\rho)$}
In this section, we will provide the proof of the inequality $\|\rho\|_{1}\leq C_{l_{1}}(\rho)$.\\
To achieve our goal, let us consider an arbitrary two-qubit quantum state described by the density operator $\rho$ in the
computational basis as
\begin{eqnarray}
\rho=
\begin{pmatrix}
  t_{11} & t_{12} & t_{13} & t_{14} \\
  t_{12}^{*} & t_{22} & t_{23} & t_{24} \\
  t_{13}^{*} & t_{23}^{*} & t_{33} & t_{34} \\
  t_{14}^{*} & t_{24}^{*} & t_{34}^{*} & t_{44}
  \end{pmatrix}, \sum_{i=1}^{4}t_{ii}=1
\label{qutrit-qubitstate}
\end{eqnarray}
where $(*)$ denotes the complex conjugate.\\\\
$l_{1}$ norm of coherence of $\rho$ is given by
\begin{eqnarray}
C_{l_{1}}(\rho)=2 \sum_{i,j=1,i\neq j}^{4} |t_{ij}|
\label{l1norm}
\end{eqnarray}
$\|\rho_{1}\|_{1}$ can be defined as
\begin{eqnarray}
\|\rho\|_{1}&=&max_{1\leq j \leq 4} \{|t_{1j}|+|t_{2j}|+|t_{3j}|+|t_{4j}|\}\nonumber\\
&=&max \{|t_{11}|+|t_{12}^{*}|+|t_{13}^{*}|+|t_{14}^{*}|, \nonumber\\&&|t_{12}|+|t_{22}|+|t_{23}^{*}|+|t_{24}^{*}|,
\nonumber\\&& |t_{13}|+|t_{23}|+|t_{33}|+|t_{34}^{*}|,\nonumber\\&& |t_{14}|+|t_{24}|+|t_{34}|+|t_{44}|\}
\label{1norm}
\end{eqnarray}
From (\ref{l1norm}) and (\ref{1norm}), it is clear that the following inequality holds
\begin{eqnarray}
\|\rho\|_{1}\leq C_{l_{1}}(\rho)
\label{reqineq}
\end{eqnarray}

\end{document}